\documentclass[a4paper,10pt]{article}
\usepackage{graphicx}

\begin{document}

\title{Use of floating surface detector stations for the calibration of a deep-sea neutrino telescope \footnote{Published in NIMA, doi:10.1016/j.nima.2008.07.011}}

\author{A.G. Tsirigotis\footnote{Physics Laboratory, School of Science and Technology, Hellenic Open University, Greece}, G. Bourlis, N.A.B. Gizani, A. Leisos \& S.E. Tzamarias}
\date{}
\maketitle

\paragraph{Abstract}
We propose the operation of floating Extensive Air Shower (EAS) detector stations in coincidence
with the KM3NeT Mediterranean deep-sea neutrino telescope to determine the absolute position and
orientation of the underwater detector and to investigate possible systematic angular errors. We
evaluate the accuracy of the proposed calibration strategies using a detailed simulation of the EAS and
KM3NeT detectors.\\


\section{Introduction}

The KM3NeT consortium is currently working on a conceptual
design for a future Mediterranean neutrino telescope, which will
have an instrumented volume of a scale of 1 km$^{3}$ [1]. A floating
array of Extensive Air Shower (EAS) detectors can be used as a seatop
calibration infrastructure, on top of the KM3NeT neutrino
telescope. Such an array can detect the copiously produced, low
energy and small zenith angle, atmospheric showers and the
collected data can be used for the reconstruction of the direction
and of the impact parameter of the shower axis. The EAS detector
array used in this study consists of floating HELYCON (HEllenic
LYceum Cosmic Observatories Network) scintillation counters [2].

According to Monte Carlo studies, presented in this paper, 35\% of
the cosmic showers in the energy range of 10$^{14}–-5\times 10^{15}$ eV contain
energetic muons able to penetrate the sea water and reach the
KM3NeT detector. These muons are detectable by the deep-sea
telescope and the muon track parameters can be estimated with
high accuracy. The comparison of the reconstructed muon track
parameters with the direction and the impact of the shower axis can
be used in order to (a) reveal systematic angular errors in the
determination of the track parameters by the neutrino telescope
and (b) provide an absolute positioning of the undersea detectors.

\section{Helycon}

\subsection{Description}

The HELYCON EAS detector array consists of detector stations
distributed over western Greece. Each station consists of charged
particle detectors, a GPS antenna, digitization and control
electronics, as well as a data acquisition system controlled by a
personal computer. The collected data are broadcasted through
the Internet to a main counting room, while the synchronization
between the HELYCON stations relies on the GPS time-signal. A
single station is able to detect atmospheric showers initiated by
cosmic particles of energy more than 10$^14$ eV.

The HELYCON charged particle detector is a scintillation
counter of 1 m$^2$ active area. It is made of plastic scintillator tiles
wrapped in Tyvek reflective paper, while the light is collected
by wave shifting fibers embedded inside the grooves of the
scintillating tiles and it is detected by a fast photomultiplier tube.

The HELYCON Readout [3] card utilizes the High Precision Time
to Digital Converter (HPTDC) chip, designed at CERN [4], and
offers up to five analog inputs, each one for a scintillation detector.
The input signals are compared to six predefined (remotely
adjustable) thresholds and the corresponding times of the PMT
waveform-threshold crossings are digitized with an accuracy of
100 ps by the HPTDC. The trigger is realized in the Field
Programmable Gate Array (FPGA) of the Readout card which is
also responsible for formatting the data and for communicating
with the station (local) PC. The data are saved on the hard disk of
the local computer and transmitted on request, via the Internet, to
a central server.

\subsection{Calibration and performance}

The charged particle detectors, before their commission,
undergo several evaluation tests and calibration procedures,
including the calibration of the photomultiplier tubes, measurements
of the response of each detector to a minimum ionizing
particle (MIP), evaluation of the uniformity of each detector
response with respect to the MIP incident point, as well as the
synchronization of all detectors belonging to the same HELYCON
station. 

The performance of HELYCON in detecting and reconstructing
showers has been studied by operating a system of eight detectors
in the laboratory. Two of the detectors were used in coincidence in
order to define an event selection trigger, while the waveforms
of the other six detectors, divided in two groups A and B, were
digitized. The collected experimental data were compared to
simulation predictions. In particular, the CORSIKA air shower
simulation software [5] was used to produce air showers, initiated
by cosmic protons entering isotropically the upper atmosphere,
whilst the detector response was simulated by the specific
HELYCON MC package. The events produced by the simulation
were stored using the same format as the experimental data and
were analyzed as real events.

A detector was considered to be active if its signal exceeded in
charge the signal corresponding to four MIPs. Subsequently, an
event was considered as a candidate of a ''shower event'' if at least
one of the detector groups A or B was active. Fig. 1 demonstrates
the success of the simulation package to describe the charge
distribution of an active detector in a shower event.

In order to evaluate the performance of a typical HELYCON
station, the analysis focused on the subset of events in which all
six detectors were active. The axis direction of every single one of
the above events was reconstructed by using (a) all six detectors,
(b) only the group of detectors A and (c) only the group of
detectors B. The direction of the shower axis was reconstructed by
using the relative arrival times of the detector signals and
implementing a simple plane-wave hypothesis for the shower
front. In the case of MC analyzed events, the possibility exists for a
direct comparison of the reconstructed to the generated zenith
angle values revealing the angular resolution of each detector setup.
The corresponding resolution values for the detector group A,
detector group B and all six detectors are

\noindent
$\sigma_{A}^{MC} = 4.5^{\circ} \pm 0.5^{\circ}, \sigma^{MC}_{B} = 5.2^{\circ} \pm 0.6^{\circ}$ \\
\noindent
$\sigma_{6}^{MC} = 3.5^{\circ} \pm 0.3^{\circ} $

\noindent
One can also evaluate the HELYCON station resolution, solely
from the real data, by comparing the results obtained by the two
detector groups (A and B) on an event–by–event basis. Fig. 2
presents the distribution of the difference ($\Delta \theta = \theta_{A} - \theta_{B}$) of these

\begin{figure}
\begin{center}
\includegraphics*[width=7cm]{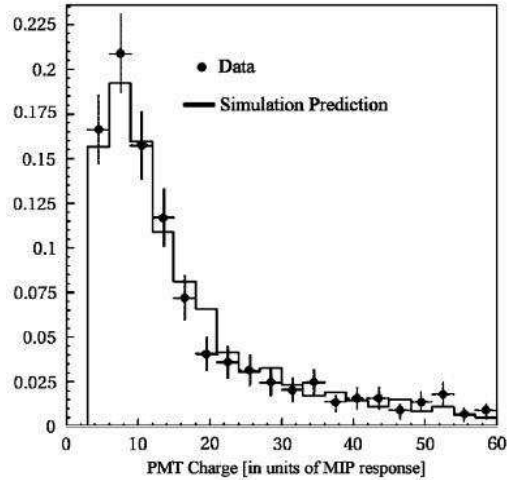}
\end{center}
\caption{The charge distribution of an active HELYCON detector in a shower event.}
\label{fig1}
\end{figure}

two estimated values of the zenith angle. The histogram
corresponds to the MC prediction while the solid curve represents
a Gaussian fit to the data points, with a sigma parameter,
$\sigma_{DATA} =7.2^{\circ} \pm 0.2^{\circ}$. This spread is consistent with the MC prediction for the resolution of each detector group 

\noindent
$\sigma^{MC} = \sqrt{(\sigma^{MC}_{A})^2 +(\sigma^{MC}_{B})^2} = 6.9^{\circ} (\pm 0.5^{\circ})$


\begin{figure}
\begin{center}
\includegraphics*[width=7cm]{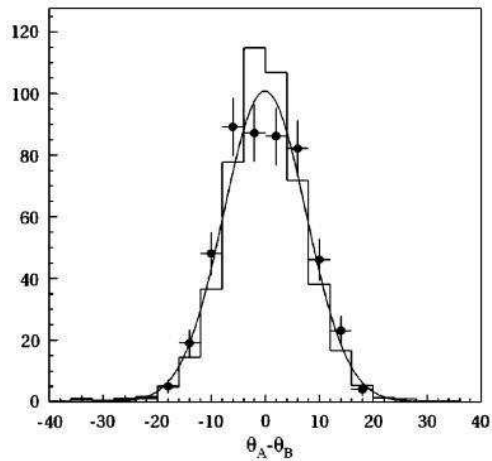}
\end{center}
\caption{The distribution of the difference of the two detector–group estimations. The solid curve represents a Gaussian fit to the data points, while the histogram corresponds to the simulation prediction.}
\label{fig2}
\end{figure}

\section{KM3NeT calibration with HELYCON detector arrays}

\subsection{Calibration set-up}
The analysis of atmospheric showers observed in coincidence
by the Antarctic Muon and Neutrino Detector Array (AMANDA) [6]
and the South Pole Air Shower Experiment (SPASE) [7] resulted in
the calibration and survey [8] of the deep AMANDA detector. In
the case of a deep-sea neutrino telescope, as the Mediterranean
KM3NeT, a cosmic shower array on the nearby shore cannot
be used for calibration purposes. However, a floating HELYCON
detector array on top of the neutrino telescope can be used to
calibrate the deep-sea detector. This array employs the copiously
produced, low energy and small zenith angle, atmospheric
showers. According to Monte Carlo studies, 35\% of the cosmic
showers in the energy range of $10^{14}-5\times 10^{15}$ eV contain energetic
muons (E$>$2 TeV) able to penetrate the 4000m deep-sea water
and reach the KM3NeT detector. Moreover, the two-thirds of these
muons can be reconstructed by the neutrino telescope and their
direction can be evaluated with an accuracy of 0.1$^{\circ}$.

The calibration capabilities of three autonomous HELYCON
detector arrays on floating platforms were quantified by a Monte
Carlo study using (a) CORSICA to simulate EAS, (b) the HELYCON
detector simulation package and (c) KM3Sim [9] a GEANT4 [10]
based simulation package to describe the passage (energy losses,
electromagnetic shower production, multiple scattering, Cherenkov
light emission) of muons through the water and the operation
of a large underwater neutrino telescope. In this study the
platforms, equipped with a dynamic positioning system, were
assumed to float 4000m above the neutrino telescope, 150m
apart of each other, around the vertical symmetry axis of the
telescope. Each platform contained 16 HELYCON detectors, arranged
on a two-dimensional grid (5m$\times$5m cell size) covering
an area of about 360 m$^{2}$. It was assumed that every single floating
detector array was operated independently from the others, as an
isolated detector.

In this study, the neutrino telescope was assumed to have
the IceCube hexagonal geometry [11], while the Optical Module
(OM) consisted of 40 cylindrical PMTs 3 in. diameter inside a 17 in.
benthos sphere, covering 4$\pi$ in solid angle [12]. This study is
based on the assumption that the relative positions of the OMs of
the KM3NeT are known by using for example acoustical positioning
techniques. The simulated response of the neutrino telescope
to down-coming muons, produced in the EAS, was used to
estimate the muon track parameters.

\subsection{Investigation for a systematic angular offset}

The simulated response of each HELYCON detector array to EAS
was analyzed in order to reconstruct the direction of the shower
axis, as described in Section 2, employing only detector signals
from the same detector station. A minimum of three active
detectors, each with a signal exceeding four MIPs, was required to
define an event as a ''shower event'' candidate. Any shower event
that contained at least one energetic muon reconstructed by the
neutrino telescope was used in the calibration and the estimated
zenith angles of the shower axis and of the muon track were
compared on an event–by–event basis. The difference between
these two angles should follow a normal distribution with mean
value equal to zero. Any statistical significant deviation of this
mean value from zero indicates that the estimations of the
neutrino telescope suffer from a systematic angular offset. The
sigma parameter of the Gaussian fit expresses the calibration
resolution per shower.

\begin{figure}
\begin{center}
\includegraphics*[width=7cm]{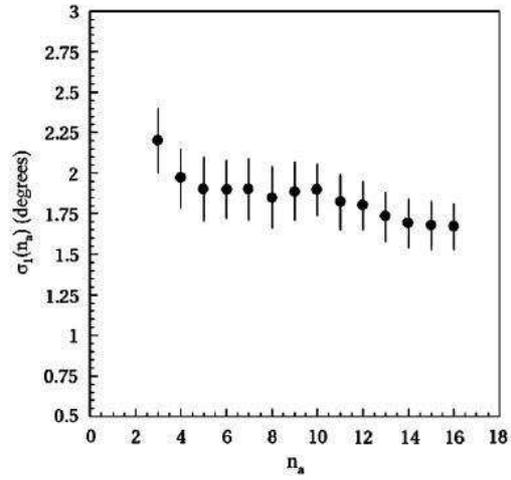}
\end{center}
\caption{The calibration resolution per single shower as a function of the number of
the minimum active detectors in the shower event.}
\label{fig3}
\end{figure}

The calibration resolution per single shower decreases when
events with more active detectors are selected, as shown in Fig. 3,
because the reconstruction accuracy of the shower's direction
improves. However, the requirement of more active detectors per
event results to a reduction in the effective area of the floating
detector array. The calibration resolution, scðnaÞ, in identifying
a possible angular offset in the neutrino telescope estimations
using the three floating detector arrays, is approximated by the
following equation:

\noindent
$\sigma_{c}(n_{a}) = \frac{\sigma_{1}(n_{a})}{\sqrt{N}}=\frac{\sigma_{1}(n_{a})}{\sqrt{3<A_{eff}(n_a)>_{E} 2\pi \int_{\Delta E} \Phi (E) dE \Delta T}} (1) $

where na is the minimum number of active detectors per shower
event, $\sigma_{1}(n_a)$ is the calibration resolution per single shower, $N$ is
the total number of shower events containing energetic muons
reconstructed by the HELYCON detector arrays and the neutrino
telescope, $<A_{eff}(n_a)>_{E} $ is the averaged over energy effective area
of one detector array, $\Phi (E)$ is the cosmic ray differential flux, $E$ is
the cosmic ray energy and $\Delta T$ is the time of operating the three
floating detector arrays. Fig. 4 shows this calibration resolution for
10 days of operation, for different event selection criteria,
demonstrating that the proposed calibration system will be able
to measure a possible zenith angle offset with an accuracy of
0.05$^{\circ}$\footnote{In this estimation the effect of the platform inclination due to winds and
waves has not been taken into account. However, this is not a problem since the
platform inclination can be accurately measured using a high precision tiltmeter
on the platform. Commercially available digital tiltmeters can offer an accuracy
much better than 0.05$^{\circ}$.}. Furthermore, shower events with less than five active
detectors do not contribute significantly in the calibration
performance of the floating arrays.

\begin{figure}
\begin{center}
\includegraphics*[width=7cm]{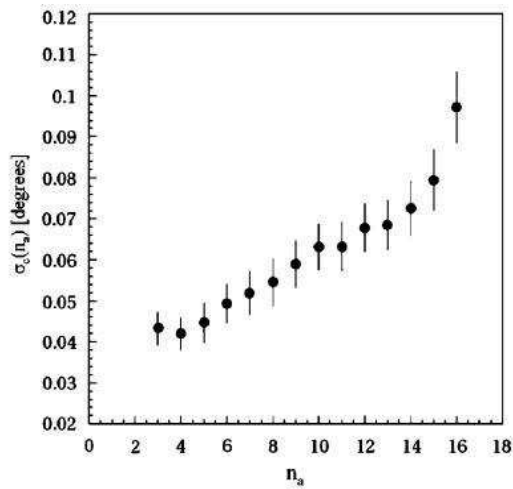}
\end{center}
\caption{The calibration resolution of three HELYCON detector arrays, for 10 days of
operation, as a function of the minimum number of active detectors per event}
\label{fig4}
\end{figure}

\subsection{Estimation of the KM3NeT absolute position}

The absolute position of the underwater telescope can be
also measured using the floating HELYCON detector arrays. The
technique is based on measuring the distance, on an event–by–
event basis, between the impact points of the reconstructed muon
track and the shower axis on the sea surface. The accuracy of this
calibration technique can be quantified by a similar expression
as Eq. (1). The position calibration resolution per single shower
ranges from 20–35m depending on the number of active
detectors. Assuming that the position of the floating array is
known with accuracy much better than this resolution,2 the
operation of three floating detector arrays, collecting data for 10
days, can provide an estimation of the absolute position of the
neutrino telescope with an accuracy of about 0.6 m.

\section{Conclusions} Detailed Monte Carlo studies, presented in this paper, have
shown that three floating HELYCON detector arrays can measure a
possible angular offset in the neutrino telescope estimations with
an accuracy of 0.05$^{\circ}$, while the absolute position of the underwater
detector can be estimated with an accuracy of about half a meter.

\section{References}
\noindent
[1] {\tt http://www.km3net.org/}. \\
\noindent
[2] {\tt http://helycon.eap.gr}; A.G. Tsirigotis, ''HELYCON: a progress report,'' In:
Online Proceedings of the 20th European Cosmic Ray Symposium (ECRS
2006) ({\tt http://www.lip.pt/events/2006/ecrs/proc/}); S.E. Tzamarias, ''HELYCON:
towards a sea-top infrastructure,'' in: Proceedings of the 6th International
Workshop on the Identification of Dark Matter (IDM 2006), World
Scientific (2007), p. 464 (ISBN-13978-981-270-852-6).\\
\noindent
[3] G. Bourlis, et al., Time over threshold electronics for an underwater neutrino
telescope, In: Proceedings of the International Workshop on a Very Large
Volume Neutrino Telescope for the Mediterranean Sea (VLVnT 2008), Nucl.
Instr. Meth. Phys. Res. to be published. \\
\noindent
[4] {\tt http://tdc.web.cern.ch/TDC/hptdc/hptdc.htm}.\\
\noindent
[5] J. Knapp, D. Heck, Nachr. Forsch. zentr. Karlsruhe 30 (1998) 27
{\tt http://www-ik.fzk.de/corsika/}.\\
\noindent
[6] {\tt http://amanda.uci.edu/}.\\
\noindent
[7] {\tt http://www.bartol.udel.edu/spase/}.\\
\noindent
[8] J. Ahrens, et al., Nucl. Instr. Meth. Phys. Res. Sect. A 522 (2004) 347.\\
\noindent
[9] A. Tsirigotis, Ph.D. Thesis, Hellenic Open University 2004. {\tt http://physicslab.eap.gr/thesis/tsirigotis}.\\
\noindent
[10] S. Agostinelli, et al., Nucl. Instr. Meth. Phys. Res. Sect. A 506 (2003) 250. \\
\noindent
[11] {\tt http://icecube.wisc.edu/}. \\
\noindent
[12] H. Lohner, et al., Talk presented at the KM3NeT CDR Workshop, NIKHEF,
Amsterdam, 12–16 November 2007.\\

\end{document}